\documentclass[11pt, a4paper]{article}
\pdfoutput=1 

\usepackage[height=8.85in,width=6.75in]{geometry}

\usepackage{graphicx,rotating}     
\usepackage[bookmarksopen,colorlinks=true,linkcolor=steelblue,
citecolor=orangered,urlcolor=darkred,linktoc=all]{hyperref}

\usepackage{amsmath,amssymb}
\usepackage{slashed}
\usepackage{bm}
\usepackage{bbm}
\usepackage{cite}
\usepackage{comment}
\usepackage[utf8]{inputenc}
\usepackage{accents}
\usepackage[footnotesize]{caption}
\usepackage{cancel}
\usepackage{cleveref}

\newcommand{\abs}[1]{\left\vert {#1} \right\vert}

\makeatletter
\@addtoreset{equation}{section}

\makeatother

\usepackage{xcolor}
\definecolor{darkred}{rgb}{0.7, 0., 0.}
\definecolor{orangered}{rgb}{1,0.27,0.}
\definecolor{steelblue}{rgb}{0.275,0.51, 0.706}
\definecolor{forestgreen}{rgb}{0.13,0.55,0.13}
\definecolor{fashionfuchsia}{rgb}{0.96, 0.0, 0.63}

\usepackage{tikz}
\usepackage{tikz-feynman}
\tikzfeynmanset{compat=1.0.0}
\pgfdeclarelayer{bg}    
\pgfsetlayers{bg,main}  

\begin{document}

\hypersetup{pageanchor=false}
\begin{titlepage}

\begin{center}

\hfill CERN-TH-2025-089\\

\vskip 0.5in

{\Huge \bfseries Spontaneous Magnetogenesis
\vspace{6.5mm}
\\ at the Electroweak Phase Transition
} \\
\vskip .8in

{\Large Valerie Domcke$^a$, Yohei Ema$^{a}$, Teerthal Patel$^{b}$}

\vskip .3in
\begin{tabular}{ll}
$^a$
&\!\!\!\!\!\emph{Theoretical Physics Department, CERN, 1211 Geneva 23, Switzerland}\\
$^b$
&\!\!\!\!\!\emph{Department of Mathematics, Vanderbilt University, Nashville, Tennessee, USA
}
\end{tabular}

\end{center}
\vskip .6in

\begin{abstract}
\noindent Spontaneous $CP$ violation during the electroweak phase transition can induce a twisting of the magnetic field configuration of Standard Model dumbbells, resulting in sizable intergalactic magnetic fields and a small baryon asymmetry, in agreement with observations. We demonstrate this by coupling the electroweak gauge group of the Standard Model to an axion-like particle with a non-vanishing velocity. Studying the resulting monopole, string and dumbbell configurations, we conclude that the helicity fraction of the magnetic fields generated at the electroweak phase transition is roughly given by the dimensionless axion velocity.
\end{abstract}

\end{titlepage}

\tableofcontents
\renewcommand{\thepage}{\arabic{page}}
\renewcommand{\thefootnote}{$\natural$\arabic{footnote}}
\setcounter{footnote}{0}
\hypersetup{pageanchor=true}

\section{Introduction}
\label{sec:introduction}

The cross-over electroweak phase transition (EWPT) in the Standard Model (SM) may source short-lived solitonic electroweak field configurations through the Kibble mechanism~\cite{Kibble:1976sj}: Magnetic monopoles, associated with the breaking of the weak $SU(2)$ gauge group, combine with $Z$-strings to form `dumbbell' configurations: a monopole and antimonopole connected by a $Z$-string, sourcing a diplolar magnetic field. In the SM, these configurations are unstable~\cite{Vachaspati:1992fi} and decay annihilating the monopoles, leaving behind magnetic fields sourced both by the dumbbells themselves~\cite{Patel:2023ybi}
as well as by Higgs gradients arising during the EWPT~\cite{Vachaspati:1991nm,Vachaspati:2020blt}. These magnetic fields are randomly orientated, preserving isotropy at large scales, and are largely erased by diffusion processes in the subsequent evolution of the Universe~\cite{Brandenburg:2017neh}.

This situation changes if the magnetic fields carry helicity~\cite{Brandenburg:2017neh}. Since in the limit of large conductivity, comoving helicity becomes a conserved quantity in the thermal plasma of the early universe, the magnetic fields are not erased by diffusion but instead their power is pushed to larger scales through an inverse cascade process. These non-linear processes have been studied in detail both analytically and numerically in the framework of magnetohydrodynamics (MHD)~\cite{Durrer:2013pga,Vachaspati:2020blt}. In this paper we propose spontaneous $CP$ violation, generated by the motion of a homogeneous axion-like field $a$, as a possible origin of helicity in the magnetic fields generated at the EWPT. We demonstrate that small axion velocities, $\dot a/(f \, T_\text{EW}) \sim 10^{-7}$ with $f$ the axion decay constant and $T_\text{EW} = 140$~GeV, suffice to generate pico-Gauss magnetic fields that are strong enough to explain the absence of secondary emission in blazar observations~\cite{
Dermer:2010mm,Taylor:2011bn,MAGIC:2022piy} while simultaneously explaining the observed baryon asymmetry of the universe through spontaneous baryogenesis~\cite{Cohen:1987vi,Cohen:1988kt}.

Axion-like background fields (axions for short in the following) arise in a number of well-motivated extensions of the SM. In string theories, they arise from the compactification of extra dimensions~\cite{Svrcek:2006yi,Arvanitaki:2009fg}. Rolling axions violate $CP$ and could hence play a role in explaining the baryon asymmetry of the universe~\cite{Sakharov:1967dj}, for example through the Affleck-Dine mechanism~\cite{Affleck:1984fy} or through spontaneous baryogenesis~\cite{Cohen:1987vi,Cohen:1988kt}. They have further been suggested to address the SM hierarchy problem through cosmological relaxation~\cite{Graham:2015cka,Hook:2016mqo}; as a dynamical field driving inflation~\cite{Freese:1990rb,Anber:2009ua,Berghaus:2019whh,Maleknejad:2011jw,Adshead:2012kp} or dark energy~\cite{Frieman:1995pm}; and as a source of dark matter in the form of ultralight dark matter~\cite{Hu:2000ke} or through the kinetic misalignment mechanism~\cite{Co:2019jts}. For the purpose of this paper, we will simply consider an axion-like background field coupling to the SM EW gauge group with non-vanishing axion velocity $\dot a$, which we take to be constant during the period of interest around the EWPT. As we will demonstrate, this spontaneous $CP$ violation is imprinted on the dumbbell configurations,  `twisting' the dipolar magnetic field configurations around the $Z$-string and the Nambu monopoles. The direction of this twist is determined by the sign of the axion velocity, resulting in a non-zero net helicity of the magnetic field configuration, which survives after the dumbbells decay.
We call this mechanism \emph{spontaneous magnetogenesis}.

To demonstrate this, we derive the equations of motion and asymptotic solutions for the field configurations of the Abrikosov-Nielsen-Olesen (ANO) string and the Nambu monopole in the presence of a non-vanishing axion velocity. Using both numerical simulations and analytical estimates, we find that in the limit of small axion velocities, the dimensionless helicity fraction of the resulting magnetic field configuration is of the order $\dot a/(f \, T_\text{EW})$. This opens up the possibility of simultaneously explaining the baryon asymmetry of the universe and generating sizable intergalactic magnetic fields (IGMFs). Notably, the generation of electromagnetic fields at the EWPT, as opposed to generating primordial hypermagnetic fields, avoids the notorious challenge of overproducing baryon number when attempting to generate sufficiently large IGMFs~\cite{Uchida:2024ude}.

The remainder of this paper is organized as follows. In Sec.~\ref{sec:string_monopole_axion},
we demonstrate that a rolling axion twists the EW dumbbell configuration by studying the equations of motion 
of the EW sector with a non-zero axion velocity.
We estimate the generation of the magnetic helicity both analytically and numerically.
In Sec.~\ref{sec:implication}, we discuss the cosmological evolution of the primordial helical magnetic field generated at the EWPT.\footnote{
	Throughout this paper, we work in Heaviside Lorentz units.
}
There we will see that the axion velocity of $\dot{a}/f T_\mathrm{EW} \sim \mathcal{O}(10^{-7})$ at the EWPT can simultaneously
explain the IGMF and the baryon asymmetry of the universe. Finally, we conclude and discuss potential future directions in Sec.~\ref{sec:conclusion}.

\section{String and monopole configurations coupled to an axion}
\label{sec:string_monopole_axion}

\begin{figure}
\centering
\includegraphics[scale=0.55]{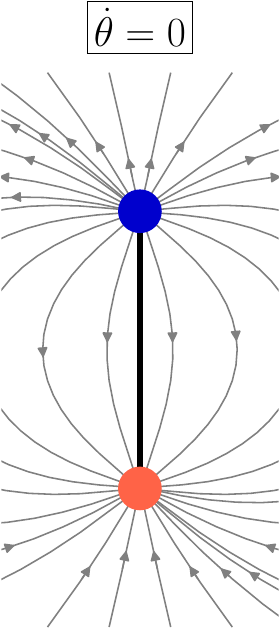}
\hspace{15mm}
\includegraphics[scale=0.55]{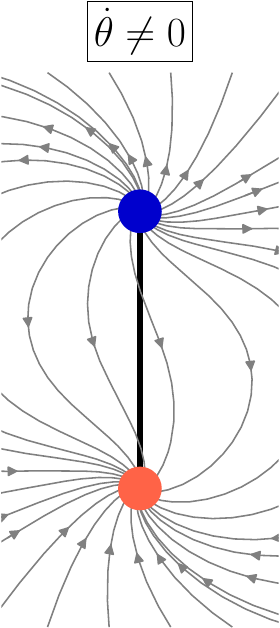}
\caption{Schematic pictures of the magnetic field configuration around the EW dumbbell without and with the axion.
The blue and red filled circles are the monopole and anti-monopole, respectively, which are connected by the $Z$-string drawn by the black solid line.
The magnetic field is a dipole configuration without the axion, which is twisted around the vertical axis ($Z$-string) by the non-zero axion velocity.
}
\label{fig:dumbbell-illustration}
\end{figure}

In this section, we study cosmic string and monopole configurations in the presence of an axion.
In general, the axion can couple to gauge bosons as
\begin{align}
	\mathcal{L}_\mathrm{int} = -\frac{a}{4f}F_{\mu\nu}\tilde{F}^{\mu\nu},
\end{align}
where $a$ is the axion field, $f$ is the axion decay constant,
and $F_{\mu\nu}$ is the field strength with $\tilde{F}^{\mu\nu} = \epsilon^{\mu\nu\rho\sigma}F_{\rho\sigma}/2$ its dual.
If the axion has a finite velocity, $\dot{a} \neq 0$, the axion introduces a chiral imbalance to the gauge sector,
leading to a twisting of the cosmic string and monopole configurations.
In particular, if the axion couples to the EW gauge bosons and is rolling during the EWPT,
the axion velocity creates a net twist of the (unstable) EW string and monopole configurations
(see Fig.~\ref{fig:dumbbell-illustration} as a schematic picture),\footnote{
	This is distinct from the twist discussed in~\cite{Vachaspati:1994ng}, which does not involve an axion.
}
which then results in a finite magnetic helicity after the decay of the EW solitons.
Intuitively, this can be understood as follows:
the rolling axion generates an effective current proportional to the magnetic field, $\vec{j} \propto \dot{a} \vec{B}$,
which then creates a magnetic field around this current, resulting in a helical magnetic field configuration.
We call this mechanism of generating the helical magnetic field from a rolling axion spontaneous magnetogenesis.

Here, we will demonstrate this by deriving the equations of motion and suitable boundary conditions for static Higgs and gauge field configurations with non-zero winding number in the presence of an axion with non-vanishing velocity, which couples to the gauge fields through the interaction above. We present analytical estimates for the total helicity of the gauge field configurations which solve these equations and confirm this by numerical simulations. We proceed in order of increasing complexity, discussing (i) cosmic strings formed in the spontaneous breaking of a $U(1)$ gauge symmetry,
(ii) cosmic strings in $SU(2) \times U(1)$ breaking, and (iii) monopoles in $SU(2) \times U(1)$ breaking. Combining the latter two leads to the unstable dumbbell configurations formed in the SM EWPT.

\subsection{Abrikosov-Nielsen-Olesen string coupled to an axion}
\label{subsec:ANO_axion}

\paragraph{Setup.} We begin with the simplest case:
the Abrikosov-Nielsen-Olesen (ANO) string~\cite{Abrikosov:1956sx,Nielsen:1973cs} in the Abelian Higgs model with an axion.
Our starting point is the action
\begin{align}
	S = \int d^4x \sqrt{-g}\left[
	\abs{D_\mu \phi}^2 - \lambda \left(\abs{\phi}^2 - \frac{v^2}{2}\right)^2 - \frac{1}{4}F_{\mu\nu}F^{\mu\nu}
	- \frac{\theta}{4\sqrt{-g}}F_{\mu\nu}\tilde{F}^{\mu\nu}\right],
\end{align}
where $\phi$ is the Abelian Higgs field,
\begin{align}
	D_\mu \phi = \left(\partial_\mu - i gA_\mu\right)\phi,
	\quad
	\tilde{F}^{\mu\nu} = \frac{1}{2}\epsilon^{\mu\nu\alpha\beta}F_{\alpha\beta},
	\quad
	\theta = \frac{a}{f},
\end{align}
with $g$ denoting the gauge coupling, 
which should not be confused with the determinant of the spacetime metric $g_{\mu\nu}$, appearing as $\sqrt{-g}$.
We fix the overall sign of the totally anti-symmetric tensor $\epsilon^{\mu\nu\alpha\beta}$ as $\epsilon^{0123} = +1$.

We assume that the axion has a constant velocity,
\begin{align}
	\dot{\theta} = \mathrm{const} \neq 0,
\end{align}
while being agnostic about its origin, and treat it as a background for the string configuration.
It is convenient to rescale the fields and spacetime variables as
\begin{align}
	\phi \to \frac{v}{\sqrt{2}}\phi,
	~~
	x^\mu \to \frac{\sqrt{2}}{gv}x^\mu,
	~~
	A_\mu \to \frac{v}{\sqrt{2}}A_\mu,
\end{align}
so that the action is given by
\begin{align}
	S &= \frac{1}{g^2}\int d^4x \sqrt{-g}\left[\abs{D_\mu \phi}^2-\frac{\beta}{2}\left(1-\abs{\phi}^2\right)^2
	-\frac{1}{4}F_{\mu\nu}F^{\mu\nu} - \frac{\theta}{4\sqrt{-g}}F_{\mu\nu}\tilde{F}^{\mu\nu}
	\right],
\end{align}
with $D_\mu \phi = (\partial_\mu - iA_\mu )\phi$ and $\beta = 2\lambda/g^2$.

\paragraph{EOMs.} 
The equations of motion are given by
\begin{align}
	0 &= \frac{1}{\sqrt{-g}}D^\mu\left(\sqrt{-g}D_\mu \phi\right) - \beta \phi \left(1-\abs{\phi}^2\right),
	\\
	0 &= \frac{1}{\sqrt{-g}}\partial_\mu\left(\sqrt{-g}F^{\mu\nu}\right)
	+ \frac{1}{\sqrt{-g}}\partial_\mu\theta\tilde{F}^{\mu\nu}
	+ i\left(\phi^* D^\nu \phi - \phi D^\nu \phi^*\right).
\end{align}
Since we are interested in the static string solution,
we use cylindrical coordinates $(t, \rho, \varphi, z)$ and assume that the solution does not depend on $t$.
We thus take the following ansatz:
\begin{align}
	\phi = e^{in\varphi} f(\rho),
	~~
	A_0 = A_\rho = 0,
	~~
	A_\varphi = n v_\varphi(\rho),
	~~
	A_z = n v_z (\rho),
\end{align}
where $\vec A = A_\rho \hat e_\rho + (A_\varphi/\rho) \hat e_\varphi + A_z \hat e_z$,
$n$ is an integer called the winding number, and we choose to work in temporal gauge.
This mimics the original ANO string configuration~\cite{Nielsen:1973cs}, 
except that we allow for a non-zero $v_z$.
This implies that we will need one additional (second order) differential equation and two boundary conditions to determine the static field configuration.
The above ansatz trivially satisfies the gauge field equations for the temporal and radial coordinate, $\nu = 0$ and $\nu = \rho$,
assuming that $f(\rho)$ is real.
The remaining non-trivial ones are given by
\begin{align}
	0 &= f'' + \frac{f'}{\rho} - \frac{n^2}{\rho^2}\left(1-v_\varphi\right)^2 f - n^2 v_z^2 f
	+ \beta f\left(1-f^2\right),
	\label{eq:string_eom_f}
	\\
	0 &= v_\varphi'' -\frac{v_\varphi'}{\rho} - \dot{\theta}\rho v_z' + 2f^2 \left(1-v_\varphi\right),
	\label{eq:string_eom_vp}
	\\
	0 &= v_z'' +\frac{v_z'}{\rho} + \frac{\dot{\theta}}{\rho}v_\varphi' - 2f^2 v_z,
	\label{eq:string_eom_vz}
\end{align}
where the prime denotes the derivative with respect to $\rho$.
The last equation indicates that, in the presence of $\dot{\theta} \neq 0$,
a non-zero $v_z \neq 0$ is necessarily generated around the string due to $v_\varphi' \neq 0$.
Since the original magnetic field of the ANO string points along the $z$-axis, i.e.\ is generated by $v_\varphi'$,
the string configuration now has a non-zero total magnetic helicity,
\begin{align}
	\mathcal{H} &\equiv \frac{g^2}{4\pi^2} \int d^3x\,
	\vec{A}\cdot \vec{B} { \;\; \to \;\;}
	\frac{n^2}{2\pi}\int_0^\infty d\rho \int dz \left(v_z' v_\varphi - v_z v_\varphi'\right) \neq 0,
	\label{eq:magnetic_helicity}
\end{align}
where the magnetic helicity is defined by the gauge fields before the rescaling, and we use the rescaled quantities in the last expression.
The quantity $\vec{A}\cdot \vec{B}$ is not locally gauge-invariant although its volume integral is.
Instead, we may define a locally gauge invariant quantity measuring the helicity in the dimensionless magnetic field as
\begin{align}
	\vec{B}\cdot\vec{\nabla}\times \vec{B} = \frac{n^2}{\rho}
	\left(v_\varphi'' v_z' - v_\varphi' v_z'' - \frac{2v_z' v_\varphi'}{\rho}\right) \neq 0,
	\label{eq:local_magnetic_helicity}
\end{align}
which is again non-zero.
In other words, the axion velocity introduces a chiral imbalance in the gauge sector,
which ``twists'' the string in a particular direction. In the small $\dot{\theta}$ limit,
the magnetic helicity is linear in $v_z$ and hence in $\dot{\theta}$, indicating that
the sign of the axion velocity determines the direction of the twist.

We may substitute the above ansatz to the action.
Since the ansatz does not depend on $t$ and $z$, we can define the string tension $T$ as $T = -d^2S/dtdz$.\footnote{
	Here we define the string tension by the action.
	Note that the action contains the Chern-Simons coupling term while the energy does not.
}
With our ansatz, the dimensionless string tension (after integrating by parts the axion term) is given by
\begin{align}
	T = \frac{2\pi}{g^2}\int_0^\infty d\rho\,\rho
	\left[f'^2 + \frac{n^2}{\rho^2}\left(1-v_\varphi\right)^2f^2
	+ n^2 v_z^2 f^2 + \frac{\beta}{2}\left(1-f^2\right)^2
	+ \frac{n^2}{2}v_z'^2 + \frac{n^2}{2\rho^2} v_\varphi'^2
	+ \frac{n^2\dot{\theta}}{2\rho}\left(v_\varphi v_z' - v_z v_\varphi' \right)
	\right].
\end{align}
The extremal condition of this tension reproduces the above equations of motion, providing a useful consistency check.

\paragraph{Boundary conditions.} To solve the static string configuration, we specify the boundary condition.
For $f$ and $v_\varphi$, we follow those of the standard ANO string, i.e.,
\begin{align}
	f(\rho = 0) = 0,
	~~
	f(\rho = \infty) = 1,
	~~
	v_\varphi(\rho = 0) = 0,
	~~
	v_\varphi(\rho = \infty) = 1.
\end{align}
This requires that the core of the string is in the symmetric phase and the gauge field is differentiable,
while the Higgs is at its potential minimum and the gauge field is a pure gauge configuration sufficiently far from the string.
For $v_z$, we require that it reduces to a pure gauge configuration at $\rho \to \infty$, i.e.,
\begin{align}
	v_z(\rho = \infty) = 0.
\end{align}
To fix the boundary condition at the core, we note that
the magnetic field is given by
\begin{align}
	\vec{B} = -\vec{\nabla}\times \vec{A} = n v_z' \hat{e}_\varphi - \frac{nv'_\varphi}{\rho}\hat{e}_z.
\end{align}
Since the $\varphi$-component is not well-defined at the origin of the core, we may require that
\begin{align}
	v_z'(\rho = 0) = 0,
\end{align}
as the final boundary condition, to ensure a differentiable field configuration at the core.
This requires that the magnetic field is straight right at the center of the core and is twisting only away from the center.
This completes the set of six boundary conditions required to fix the solutions of the three second order differential equations above.

\begin{figure}[t]
	\centering
 	\includegraphics[width=0.4\linewidth]{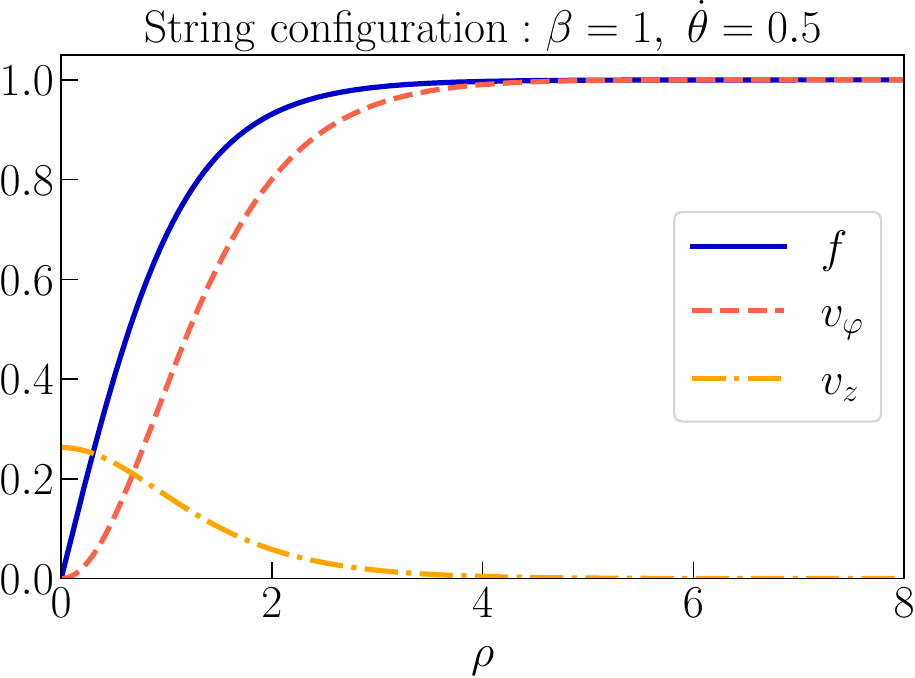}
	\hspace{2.5mm}
	\vspace{1.5mm}
	\includegraphics[width=0.4\linewidth]{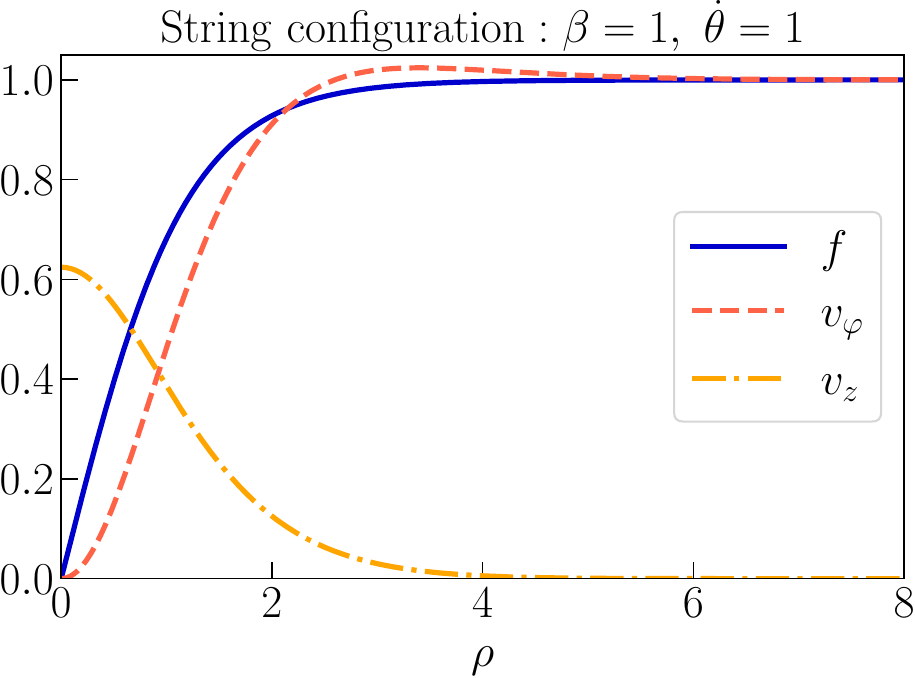}
 	\includegraphics[width=0.4\linewidth]{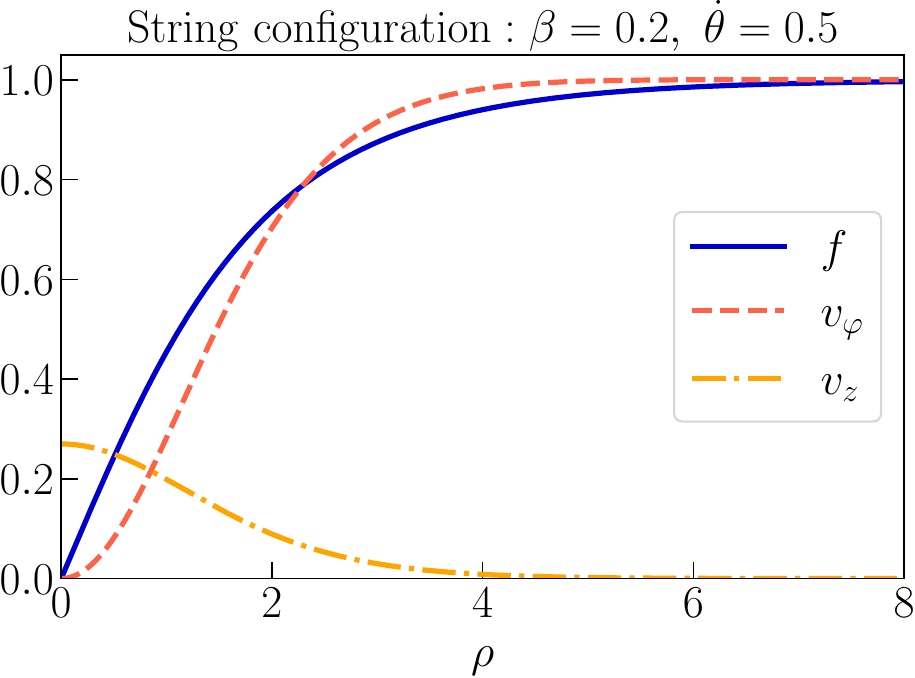}
	\hspace{2.5mm}
	\includegraphics[width=0.4\linewidth]{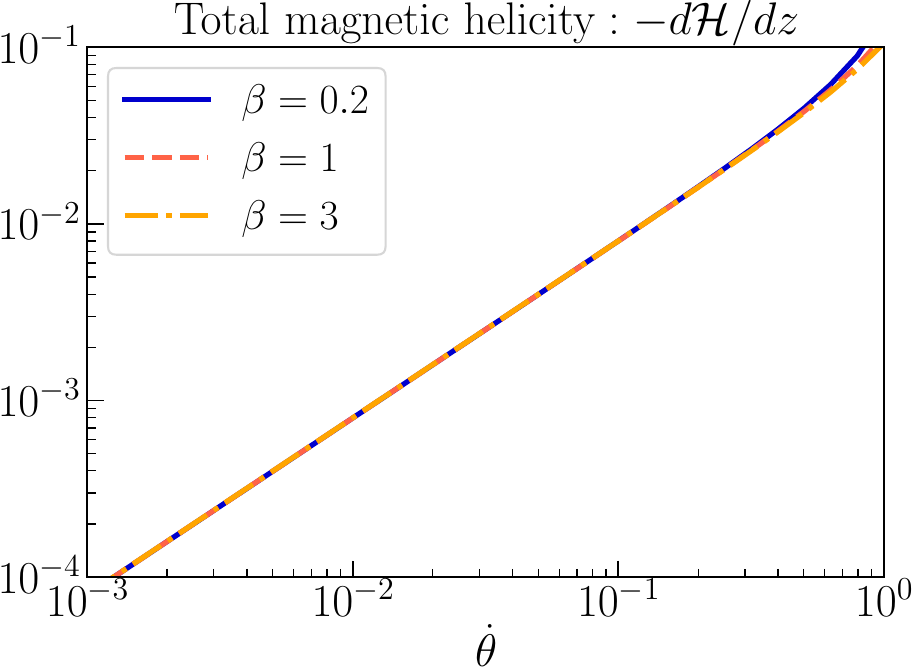}
	\caption{\small \emph{Upper and lower-left panels}: the ANO string configuration in the presence of an axion velocity
	for $\dot{\theta} = \{ 0.5, 1\}$ and $\beta = \{0.2, 1\}$ in rescaled units and setting $n = 1$.
	The non-zero axion velocity induces the $\varphi$ component of the gauge field,
	resulting in a net magnetic helicity of the configuration.
	\emph{Lower-right panel}: the magnetic helicity as defined in Eq.~\eqref{eq:magnetic_helicity}
	per unit length of the string as a function of the axion velocity for $n = 1$ and three different values of $\beta$.
	This shows that the magnetic helicity is linear in the axion velocity,
	and is almost insensitive to $\beta$, for $\dot{\theta} \ll 1$.
	}
	\label{fig:NO_axion}
\end{figure}

\paragraph{Numerical simulation.} In the top and bottom-left panels of Fig.~\ref{fig:NO_axion}, 
we plot the string configurations for $n = 1$ obtained numerically by the relaxation method.
Specifically, we follow the gradient flow method in~\cite{Eto:2022hyt} 
and introduce a fictitious ``flow time'' $\tau$ to the Higgs and gauge bosons.
We then solve the differential equations after adding $\partial_\tau f, \partial_\tau v_\varphi$, 
and $\partial_\tau v_z$ on the left-hand sides of Eqs.~\eqref{eq:string_eom_f}\,--\,\eqref{eq:string_eom_vz}, 
respectively.
Once we take the time $\tau_f$ at which we evaluate the configuration to be sufficiently large,
the solution does not depend on $\tau_f$ anymore and is static, providing the desired string configuration.
As one can see from the plots, the non-zero axion velocity necessarily induces $v_z$ as we argued above,
and the magnitude of $v_z$ increases with~$\dot{\theta}$, indicating the generation of the magnetic helicity.\footnote{
Here we focus on values of $\dot \theta \ll 1$, which will be the regime of interest for magnetogenesis and baryogenesis. For larger values of $\dot \theta$, the axion gauge-field system is known to exhibit a tachyonic instability, leading to rapid helical gauge field production. As we discuss in Sec.~\ref{sec:implication}, such large values of $\dot \theta$ generically lead to an overproduction of the observed baryon asymmetry, and hence we do not consider them any further here.
}
To see this point in more detail,
in the bottom-right panel of Fig.~\ref{fig:NO_axion},
we plot the magnetic helicity~\eqref{eq:magnetic_helicity} per unit length 
as a function of the axion velocity for $n = 1$ and several choices of $\beta$.
The figure demonstrates that the magnetic helicity is indeed generated, and its size is linear in $\dot{\theta}$ for $\dot{\theta}\ll 1$.
Moreover, we find that the magnetic helicity~\eqref{eq:magnetic_helicity} depends only very mildly
on $\beta$ for $\dot{\theta} \ll 1$, and is given by
\begin{align}
	\frac{d\mathcal{H}}{dz} \simeq -0.08 \times \dot{\theta},
\end{align}
in the rescaled units. Thus, we have checked numerically that
the axion velocity indeed twists the magnetic field around the ANO string.
Note that flipping the sign of $n$ does not change the sign of the magnetic helicity, as the latter is proportional to $n^2$;
the sign of the magnetic helicity is controlled purely by the axion velocity.

\paragraph{Semilocal strings.} The above analysis can be trivially extended to semilocal strings~\cite{Vachaspati:1991dz,Achucarro:1999it}.
To see this, let us consider the following action
\begin{align}
	S = \int d^4x \sqrt{-g}\left[
	\abs{D_\mu \Phi}^2 - \lambda \left(\abs{\Phi}^2 - \frac{v^2}{2}\right)^2 - \frac{1}{4}Y_{\mu\nu}Y^{\mu\nu}
	- \frac{\theta}{4\sqrt{-g}}Y_{\mu\nu}\tilde{Y}^{\mu\nu}\right],
\end{align}
where $\Phi$ is now an $SU(2)$ doublet and
we denote the abelian gauge field as $Y_\mu$ with its field strength $Y_{\mu\nu}$, 
anticipating that it is identified with the SM hypercharge $U(1)$ field.
The covariant derivative is defined as
\begin{align}
	D_\mu \Phi = \left(\partial_\mu - i \frac{g'}{2}Y_\mu\right)\Phi.
\end{align}
This model corresponds to the $g \to 0$ (or $\sin\theta_W \to 1$) limit of the SM.
It possesses a global $SU(2)$ symmetry and a local $U(1)$ symmetry,
and after the symmetry breaking there remains a global $U(1)$ symmetry.
If we look at the entire (global) $\times$ (local) symmetry, it does not have a non-trivial fundamental group,
but if we focus on the gauge sector only, the symmetry breaking pattern is identical to the Abelian Higgs model,
indicating that there is a non-trivial configuration, i.e.\ a cosmic string.
Recall that the isolated global string has infinite energy due to the gradient term of the Higgs,
while the gauge field compensates for this in the case of the local string.
If the Higgs field spans the gauged $U(1)$ orbit of the vacuum manifold initially, to shrink this circle to a point,
we need to go outside of the gauged $U(1)$ orbit and enter the region spanned by the global $SU(2)$ symmetry in the vacuum manifold.
This costs additional energy due to the non-compensated gradient term,
indicating a certain stability of the initial configuration.
This is the so-called semilocal string~\cite{Vachaspati:1991dz,Achucarro:1999it}.
To study the semilocal string configuration, we again redefine the fields as
\begin{align}
	\Phi \to \frac{v}{\sqrt{2}}\Phi,
	~~
	x^\mu \to \frac{2\sqrt{2}}{g'v}x^\mu,
	~~
	Y_\mu \to \frac{v}{\sqrt{2}}Y_\mu,
\end{align}
to obtain
\begin{align}
	S &= \frac{4}{g'^2}\int d^4x \sqrt{-g}\left[\abs{D_\mu \Phi}^2-\frac{\beta}{2}\left(1-\abs{\Phi}^2\right)^2
	-\frac{1}{4}Y_{\mu\nu}Y^{\mu\nu} - \frac{\theta}{4\sqrt{-g}}Y_{\mu\nu}\tilde{Y}^{\mu\nu}
	\right],
\end{align}
where $\beta = 8\lambda/g'^2$.
The equations of motion are derived as
\begin{align}
	0 &= \frac{1}{\sqrt{-g}}D^\mu\left(\sqrt{-g}D_\mu \Phi\right) - \beta \Phi \left(1-\abs{\Phi}^2\right),
	\\
	0 &= \frac{1}{\sqrt{-g}}\partial_\mu\left(\sqrt{-g}Y^{\mu\nu}\right)
	+ \frac{1}{\sqrt{-g}}\left(\partial_\mu\theta\right)\tilde{Y}^{\mu\nu}
	+ i\left(\Phi^\dagger D^\nu \Phi - (D^\nu \Phi)^\dagger \Phi \right).
\end{align}
They take the same form as the Abelian Higgs model except that now the Higgs is a doublet.
Indeed, if $\phi_\mathrm{ANO}$ and $Y^\mu_\mathrm{ANO}$ are the ANO string configuration
of the Abelian Higgs model, the configuration
\begin{align}
	\Phi = \phi_\mathrm{ANO} \Phi_0,
	\quad
	Y^\mu = Y_\mathrm{ANO}^\mu,
\end{align}
satisfies the equation of motion of the present model, where $\Phi_0$ is a constant $SU(2)$ doublet satisfying
\begin{align}
	\Phi_0^\dagger \Phi_0 = 1.
\end{align}
This is true independent of whether there is a coupling to an axion or not.

The stability of the semilocal string configuration depends on $\beta$ and is non-trivial as it is not protected by the topology.
Although it is beyond the scope of our paper to study the stability with an arbitrary size of the axion velocity,
we do not expect a drastic change in the (in)stability of the configuration 
as long as $\dot{\theta}/v \ll 1$, which is the case of our interest later.
For a more comprehensive discussion on the stability of the semilocal strings (without the axion), 
see~\cite{Achucarro:1999it} and references therein.
In summary, we thus expect that the discussion of the local ANO string above qualitatively also applies to semilocal strings.

\subsection{Electroweak string coupled to an axion}
\label{subsec:EW_axion}

We move on to the EW sector of the SM, now including the non-zero $SU(2)$ coupling $g$. The action is given by
\begin{align}
	S = \int d^4x \sqrt{-g} & \left[
	\abs{D_\mu \Phi}^2 - \lambda \left(\abs{\Phi}^2 - \frac{v^2}{2}\right)^2 \right. \nonumber \\
		& \left. -\frac{1}{4}W_{\mu\nu}^{a}W^{a\mu\nu} - \frac{1}{4}Y_{\mu\nu}Y^{\mu\nu}
	- \frac{\theta_2}{4\sqrt{-g}}W_{\mu\nu}^{a}\tilde{W}^{a\mu\nu}
	- \frac{\theta_1}{4\sqrt{-g}}Y_{\mu\nu}\tilde{Y}^{\mu\nu}\right],
\end{align}
where the covariant derivative is given by
\begin{align}
	D_\mu \Phi = \left(\partial_\mu - i\frac{g}{2}W_\mu^a \tau^a - i \frac{g'}{2}Y_\mu\right)\Phi,
\end{align}
with $\tau^a$ the Pauli matrices. We denote $\theta_2 = g_{aWW}a/f$ and $\theta_1 = g_{aYY}a/f$.
The following discussion is analogous to~\cite{Vachaspati:1992fi}, but now with the axion coupling.
We take the ansatz
\begin{align}
	\Phi = \begin{pmatrix}
	0 \\ \phi
	\end{pmatrix},
	\quad
	W_\mu^\pm = 0,
	\quad
	\begin{pmatrix}
	Z_\mu \\ A_\mu
	\end{pmatrix}
	= \begin{pmatrix} \cos\theta_W & -\sin\theta_W \\ \sin\theta_W & \cos\theta_W \end{pmatrix}
	\begin{pmatrix} W_\mu^3 \\ Y_\mu \end{pmatrix},
\end{align}
where
\begin{align}
	\cos\theta_W = \frac{g}{\sqrt{g^2 + g'^2}},
	\quad
	\sin\theta_W = \frac{g'}{\sqrt{g^2 + g'^2}}.
\end{align}
The electromagnetic charge conservation guarantees that $W_\mu^\pm$ always appear quadratically in the action,
and hence we can consistently set them to be zero in our ansatz.
With this ansatz, the action is given by
\begin{align}
	S = \int d^4x \sqrt{-g}&\left[
	\abs{D_\mu \phi}^2 - \lambda\left(\abs{\phi}^2 - \frac{v^2}{2}\right)^2
	\right. \nonumber \\ &\left.
	-\frac{1}{4}Z_{\mu\nu}Z^{\mu\nu} - \frac{1}{4}F_{\mu\nu}F^{\mu\nu}
	-\frac{1}{4\sqrt{-g}}\left(
	\theta_Z Z_{\mu\nu}\tilde{Z}^{\mu\nu}
	+ \theta_A F_{\mu\nu}\tilde{F}^{\mu\nu}
	- 2\theta_{ZA} Z_{\mu\nu}\tilde{F}^{\mu\nu}
	\right)
	\right],
\end{align}
where we define
\begin{align}
	D_\mu \phi = \partial_\mu \phi + \frac{i\sqrt{g^2 + g'^2}}{2}Z_\mu \phi,
\end{align}
and
\begin{align}
	\theta_Z = \theta_2 \cos^2\theta_W + \theta_1 \sin^2\theta_W,
	\quad
	\theta_A = \theta_2 \sin^2\theta_W + \theta_1 \cos^2\theta_W,
	\quad
	\theta_{ZA} =  \left(\theta_1 - \theta_2\right) \sin\theta_{W} \cos\theta_W.
\end{align}
In the above, $Z_{\mu\nu}$ and $F_{\mu\nu}$ are the field strengths
constructed from $Z_\mu$ and $A_\mu$, respectively.
If the terms in the round brackets on the second line are absent, or if there is no axion in general,
the system reduces to the ANO string of the Abelian Higgs model.
The ANO string configuration is the exact solution of the equations of motion, and this configuration is 
the so-called $Z$-string~\cite{Vachaspati:1992fi,James:1992wb},
which is however known to be unstable with the measured values of the SM parameters.
The last term indicates the mixing between $Z_\mu$ and $A_\mu$ in the presence of the axion
for $\theta_2 \neq \theta_1$.
To avoid the complexity associated with the mixing, in this paper, we focus on the case of
\begin{align}
	\theta_2 = \theta_1 = \theta,
\end{align}
and leave the more general case of $\theta_2 \neq \theta_1$ for future work.\footnote{
	Concerning the EW strings, the general case $\theta_2 \neq \theta_1$ does not introduce a huge complexity;
	we can include $A_\mu \neq 0$ and derive a consistent set of static equations, which can be solved (at least numerically),
	in the same way as the original ANO string.
	The true complexity arises when we discuss the Nambu monopole in the next subsection (see~\cref{fn:monopole_general}).
}
We can then set $A_\mu = 0$ consistently (implying that the string is a pure $Z$-string and the electromagnetic field is sourced only by the monopoles).
After rescaling the fields as
\begin{align}
	\phi \to \frac{v}{\sqrt{2}}\phi,
	\quad
	x^\mu \to \frac{2\sqrt{2}/v}{\sqrt{g^2 + g'^2}}x^\mu,
	\quad
	Z_\mu \to -\frac{v}{\sqrt{2}}Z_\mu,
\end{align}
where the minus sign of $Z_\mu$ is set to match with Sec.~\ref{subsec:ANO_axion},
the action is now brought to the form
\begin{align}
	S = \frac{4}{g^2 + g'^2}\int d^4x \sqrt{-g}&
	\left[
	\abs{(\partial_\mu - i Z_\mu) \phi}^2 - \frac{\beta}{2}\left(1-\abs{\phi}^2\right)^2
	-\frac{1}{4}Z_{\mu\nu}Z^{\mu\nu}
	-\frac{\theta}{4\sqrt{-g}} Z_{\mu\nu}\tilde{Z}^{\mu\nu}
	\right],
	\label{eq:action_EW_dimensionless}
\end{align}
where $\beta = 8\lambda/(g^2 + g'^2)$.
This is the ANO string action including an axion, and therefore the discussion in Sec.~\ref{subsec:ANO_axion}
directly applies to the EW theory.
In particular, the axion induces a net $Z$-magnetic helicity.

\subsection{Nambu monopole coupled to an axion}
\label{subsec:nambu_monopole}

So far we have discussed cosmic strings in the presence of an axion.
In the case of the $Z$-string, the strings form dumbbell-type configurations with Nambu monopoles attached
at the endpoints~\cite{Nambu:1977ag}. Therefore we need to understand the effect of the axion on the Nambu monopole.
For this we follow the arguments in~\cite{Nambu:1977ag}, but including the axion.

\paragraph{Asymptotic solution.} We focus on the asymptotic solution at large distances from the soliton in this subsection. In order for the configuration to be of finite energy, it must satisfy
\begin{align}
	D_\mu \Phi = \left[\partial_\mu - i\frac{g}{2}W_{\mu}^a \tau^a - i\frac{g'}{2}Y_\mu\right]\Phi = 0,
	\quad
	V(\phi) = \mathrm{min.},
	\label{eq:monopole_asymptotics}
\end{align}
at a large distance from the monopole.
The first condition can be solved for the gauge bosons as
\begin{align}
	g W_\mu^a + g'Y_\mu \Phi^\dagger \tau^a \Phi = -i\Phi^\dagger \tau^a \overset{\leftrightarrow}{\partial}_\mu \Phi,
\end{align}
where we set $\Phi^\dagger \Phi = v^2 = 1$.
After using the Fierz identity for the $SU(2)$ indices
and decomposing to the $\Phi^\dagger \tau^a \Phi$ and its orthogonal directions,
we obtain
\begin{align}
	&gW_\mu^a = -\epsilon^{abc}\left(\Phi^\dagger \tau^b \Phi\right) \partial_\mu \left(\Phi^\dagger \tau^c \Phi\right)
	-\left(\xi i \Phi^\dagger \overset{\leftrightarrow}{\partial}_\mu \Phi + a_\mu\right) \Phi^\dagger \tau^a \Phi,
	\label{eq:W_monopole}
	\\
	&g'Y_\mu = -(1-\xi) i \Phi^\dagger \overset{\leftrightarrow}{\partial}_\mu \Phi + a_\mu,
	\label{eq:Y_monopole}
\end{align}
where $\xi$ is a number related to the weak mixing angle to be determined later,
and $a_\mu$ is an ``external'' gauge field induced by the axion velocity which we discuss below.
Correspondingly, we obtain
\begin{align}
	gW_{\mu\nu}^a
	= -\left[\left(1-\xi\right)f_{\mu\nu} + a_{\mu\nu}\right]\Phi^\dagger \tau^a \Phi,
	\quad
	g' Y_{\mu\nu} = (1-\xi) f_{\mu\nu} + a_{\mu\nu},
	\label{eq:Ysol}
\end{align}
where
\begin{align}
	f_{\mu\nu} = -2i \left[\partial_\mu \Phi^\dagger \partial_\nu \Phi - \partial_\nu \Phi^\dagger \partial_\mu \Phi\right], 
	\quad a_{\mu\nu} = \partial_\mu a_\nu - \partial_\nu a_\mu.
\end{align}
We then define the usual linear combination of $Y_{\mu\nu}$ and $W_{\mu\nu}^a \Phi^\dagger \tau^a \Phi$ as
the $Z$-boson and photon.
The field strength of $Z$-boson is given by
\begin{align}
	Z_{\mu\nu} = \frac{g}{\sqrt{g^2 + g'^2} }W_{\mu\nu}^a \Phi^\dagger \tau^a \Phi + \frac{g'}{\sqrt{g^2 + g'^2} } Y_{\mu\nu} = 0,
\end{align}
while the field strength of the photon is given by\footnote{
	Here we use the definition in~\cite{Nambu:1977ag}.
	One may instead define the electromagnetic field by including the Higgs gradient term 
	to eliminate the gauge boson self-interaction term contained in $W_{\mu\nu}^a$~\cite{tHooft:1974kcl,Vachaspati:1991nm}.
	This choice does not affect our conclusion below that the axion induces 
	the magnetic helicity around the magnetic monopole as it is derived from solving the equations of motion
	of the original $SU(2)$ and $U(1)_Y$ fields, which is independent of how we define the photon field.
}
\begin{align}
	F_{\mu\nu} = -\frac{g'}{\sqrt{g^2 + g'^2} } W_{\mu\nu}^a\Phi^\dagger \tau^a \Phi + \frac{g}{\sqrt{g^2 + g'^2} } Y_{\mu\nu}
	= \frac{\sqrt{g^2 + g'^2} }{g g'} \left[(1-\xi)f_{\mu\nu} + a_{\mu\nu}\right].
\end{align}
Therefore, well outside of the monopole, only the electromagnetic field survives.
Note that the discussion so far depends solely on Eq.~\eqref{eq:monopole_asymptotics}
and does not explicitly depend on the axion. The electromagnetic field strength does however receive a contribution from the Higgs gradients, $f_{\mu\nu}$, as well as from the field $a_\mu$ introduced in Eq.~\eqref{eq:Ysol}. As we will see below, it is exclusively the latter term that is sourced by the axion velocity.

\paragraph{Nambu monopole and Z-string.} We specify the Higgs configuration as
\begin{align}
	\Phi = \begin{pmatrix} \cos \vartheta/2 \\ e^{i\varphi} \sin \vartheta/2 \end{pmatrix}.
\end{align}
with $\varphi$ the azimuthal and $\vartheta$ the polar angle with respect to the $z$-axis,
assuming that the axion does not modify the Higgs configuration significantly.
This configuration depends on $\varphi$ even at $\vartheta = \pi$ where $\varphi$ is ill-defined, meaning that we need a string
attached to this monopole so that $\Phi$ vanishes there, which is the $Z$-string.
Note that in contrast to the 't~Hooft--Polyakov monopole~\cite{tHooft:1974kcl,Polyakov:1974ek} 
where the Higgs field is a triplet of $SU(2)$ (or a vector of $SO(3)$),
the Higgs field is a doublet in the SM, making it impossible to
find a non-trivial mapping between a 2-dimensional sphere in space and the Higgs field configuration
without this singularity.
It follows that
\begin{align}
	\Phi^\dagger \tau^a \Phi = \frac{x^a}{r},
\end{align}
and therefore, the field strengths are well-defined even at $\vartheta = \pi$,
apart from the monopole core.
By inserting this expression, we obtain
\begin{align}
	gW_{i}^a = \epsilon_{aij} \frac{x^j}{r^2}+ \left[\xi (1-\cos\vartheta) \partial_i \varphi - a_i\right]\frac{x^a}{r},
	\quad
	g'Y_i = (1-\xi) (1-\cos\vartheta)\partial_i \varphi + a_i,
	\label{eq:gauge_fields_monopole}
\end{align}
and
\begin{align}
	gW_{ij}^a = -(1-\xi)\epsilon_{bij}\frac{x^a x^b}{r^4} - a_{ij}\frac{x^a}{r},
	\quad
	g'Y_{ij} = (1-\xi)\epsilon_{bij}\frac{x^b}{r^3} + a_{ij},
\end{align}
where $\epsilon_{123} = 1$ and we set $a_0 = \partial_0 a_i = 0$.
The sum is taken for the adjoint $SU(2)$ indices when they appear twice with respect to $\delta_{ab}$,
and we do not distinguish the position of the adjoint $SU(2)$ index.
The first term indeed provides the magnetic monopole configuration.

To determine $\xi$, we assume that the $Z$-string attached to the monopole 
does not contain any electromagnetic field flux~\cite{Nambu:1977ag,Vachaspati:1994xe}.
The hypermagnetic field flux emitted from the monopole, excluding the contribution from the singular line $\vartheta = \pi$, is given by
\begin{align}
	g' \int_{\vartheta \neq \pi} d\vec{S} \cdot \vec{B}_Y = 4\pi(1-\xi),
\end{align}
while the one flowing into the monopole at the $Z$-string at $\vartheta = \pi$ is
\begin{align}
	g' \int_{\vartheta = \pi} d\vec{S} \cdot \vec{B}_Y = 4\pi (1-\xi),
\end{align}
where we used Eq.~\eqref{eq:gauge_fields_monopole} and Stokes' theorem.
Note that these two contributions are equal, as they should be, due to the $U(1)$ nature of the hypercharge.
On the other hand, the $Z$-flux coming into the monopole from the $Z$-string is quantized as $2\pi n$ in terms of the rescaled fields.
Recalling that the relevant coupling for the $Z$-string is $\sqrt{g^2 + g'^2}/2$ as given in Eq.~\eqref{eq:action_EW_dimensionless},
we obtain for the $Z$-string with the winding number $n = 1$
\begin{align}
	\sqrt{g^2 + g'^2}\int_{\vartheta = \pi} d\vec{S} \cdot \vec{B}_Z = 4\pi.
\end{align}
Therefore, the magnetic flux in the $Z$-string is given by
\begin{align}
	\int_{\vartheta = \pi} d\vec{S} \cdot \vec{B} = 
	\int_{\vartheta = \pi} d\vec{S} \cdot \left[\frac{g'}{\sqrt{g^2 + g'^2}}\left(\frac{g'}{g}+\frac{g}{g'}\right)\vec{B}_Y-\frac{g'}{g}\vec{B}_Z\right]
	= \frac{4\pi}{\sqrt{g^2 + g'^2}}\left[\frac{g}{g'}(1-\xi) - \frac{g'}{g}\xi\right].
\end{align}
By setting it to zero, as this minimizes the energy of the $Z$-string, we obtain
\begin{align}
	\xi = \frac{g^2}{g^2 + g'^2} = \cos^2\theta_W.
\end{align}

\paragraph{Magnetic helicity.} The discussion so far relies solely on the finite energy condition in the Higgs sector, Eq.~\eqref{eq:monopole_asymptotics},
but the gauge fields should moreover obey their equations of motion.
In the case of $\theta_2 = \theta_1 = \theta$,
the equations of motion of the $SU(2)$ and $U(1)$ gauge fields are identitcal and are given by\footnote{
	For $\theta_2 \neq \theta_1$, the equations are not identical, and therefore cannot be solved
	solely by $a_\mu$.
	In this case, we might need to modify the condition~\eqref{eq:monopole_asymptotics}
	as the mixing between the photon and $Z$-boson might induce a non-zero $Z$ configuration,
	which then may lead to a deviation from Eq.~\eqref{eq:monopole_asymptotics}.
	\label{fn:monopole_general}
}
\begin{align}
	0 = \partial_\mu \left[\sin^2 \theta_Wf^{\mu\nu} + a^{\mu\nu}\right] + \partial_\mu \theta \left[\sin^2 \theta_W\tilde{f}^{\mu\nu} + \tilde{a}^{\mu\nu}\right].
\end{align}
For a static gauge field configuration in temporal gauge, $a_0 = \partial_0 a_i = 0$,
and inserting the explicit configuration of the Higgs field, this reduces to
\begin{align}
	0 &= \partial_j a_{ji} + \frac{\dot{\theta}}{2}\epsilon^{ijk}a_{jk} + \dot{\theta} \sin^2 \theta_W\frac{x^i}{r^3}.
	\label{eq:eom_monopole_red}
\end{align}
By solving this equation, one could obtain the full magnetic field configuration of the Nambu monopole with the axion.

Instead of solving the full equation,
here we focus on the net magnetic helicity indicated by this equation.
By noting that $B_i = -\epsilon^{ijk}F_{jk}/2$, we may define its two contributions as
\begin{align}
	B^i = B^i_a + B^i_f,
	\quad
	B^i_a = \frac{1}{2}\epsilon^{ijk}a_{jk},
	\quad
	B_f^i = \frac{\sin^2 \theta_W}{2}\epsilon^{ijk}f_{jk} = \sin^2 \theta_W\frac{x^i}{r^3},
\end{align}
where $B_a$ will ultimately be sourced by the axion and $B_f$ is the SM contribution associated with the gradients of the Higgs field.
Ignoring the second term which is suppressed by $\dot \theta a_{\mu\nu} \sim \dot{\theta}^2$, Eq.~\eqref{eq:eom_monopole_red} tells us that
\begin{align}
	0 \simeq \vec{\nabla}\times \vec{B}_a - \dot{\theta} \vec{{B}_f},
\end{align}
from which one obtains the local magnetic helicity as
\begin{align}
	\vec{B}\cdot \vec{\nabla}\times \vec{B}
	\simeq \dot{\theta} {B}_f^2.
	\label{eq:h_estimate}
\end{align}
Therefore, there is a net magnetic helicity proportional to the axion velocity, 
and the magnitude of this (dimensionless) helicity is set roughly by $\dot \theta$. 
Since it depends quadratically on $\vec{B}_f$, the induced magnetic helicity has a definite sign fixed by the sign of the axion velocity,
independent of the sign of the monopole charge.

We comment on a caveat of the above argument.
In the absence of SM fermions, a non-zero $\theta_2$ transforms the Nambu monopole to a dyon
due to the Witten effect~\cite{Witten:1979ey,Vachaspati:1994xe},
indicating that the monopole obtains the electric charge due to the axion motion, developing non-zero $a_0$.
This is compensated by chiral fermion charge generation
in the presence of the SM fermions~\cite{Callan:1982ah,Callan:1982au,Rubakov:1982fp,Hamada:2022bzn} 
as one can rotate away a constant $\theta_2$ by the chiral transformations, making it unphysical.
We have ignored these potentially interesting and rich dynamics of the SM fermions in the presence of the Nambu monopole with the axion,
and instead focused on one particular static solution of the equations of motion.
We leave a more systematic study on the role of SM fermions in this context for future work.\footnote{
    Our choice $\theta_1 = \theta_2$ is for simplicity. We expect that the SM fermions are less relevant for the choice of $\theta_2 = 0$.
}

\paragraph{Chiral asymmetry in SM fermions.}
Before closing this subsection, we comment on the possible impact of primordial chiral chemical potentials associated with the SM fermions.
Once the fermions are included, we can perform a chiral rotation to eliminate the axion-gauge field coupling, at the price of coupling the axion instead derivatively to the axial vector currents, mimicking the chiral chemical potential.
Therefore, it is natural to ask if a primordial SM fermion asymmetry can twist the EW solitons even without the presence of a rolling axion.

Although the chiral chemical potential in general plays an identical role to the axion 
through the chiral magnetic effect (CME), unfortunately, the answer to this question is negative within the SM.
In the SM, all the interactions are in thermal equilibrium at the EWPT, and only the chemical potentials
associated with the conserved charges, $U(1)_{B-L}$, $U(1)_{L_i - L_j}$, and $U(1)_Y$, can be non-vanishing (depending on the initial condition).
These conserved charges are anomaly-free by definition and therefore do not introduce a chiral imbalance in the EW sector.
Thus, we need an additional source of the chiral imbalance, such as the axion field.
This should not be a surprise since the CME in general requires out-of-thermal equilibrium environment.
We discuss this point in more detail in App.~\ref{app:cme_fermion} in a field-redefinition-independent way.

\section{Implications for intergalactic magnetic field}
\label{sec:implication}

We have seen in the previous section that the EW solitons acquire ``twists'' in the presence of the axion coupling to the EW gauge bosons.
As a result, the magnetic field generated after the decay of the EW solitons at the EWPT obtains non-zero net magnetic helicity.
The presence of this net helicity, a conserved quantity within the SM plasma, alters the subsequent evolution of cosmological magnetic fields, enhancing their magnitude today.
In particular, we will see that a modest value of the axion velocity at the EWPT can source a relatively strong intergalactic magnetic field, as motivated by blazar observations~\cite{Dermer:2010mm,Taylor:2011bn,MAGIC:2022piy}.

\subsection{Helical magnetic field production at EWPT}
\label{subsec:helical_B_EWPT}

The formation of magnetic fields during electroweak symmetry breaking in the SM is a challenging problem. Ref.~\cite{Vachaspati:1991nm} pointed out the possibility of temporarily forming magnetic fields associated with gradients of the Higgs field and estimated the correlation length to be set by $(g T_\text{EW})^{-1}$, see also discussion in \cite{Durrer:2013pga}. However, classical lattice gauge field simulations of the $SU(2) \times U(1) \mapsto U(1)$ breaking process indicated larger correlations lengths~\cite{Diaz-Gil:2008raf,Zhang:2019vsb}.
Arguments have been put forward that the relevant scale may in fact be the Hubble scale at the EWPT: From the Kibble mechanism~\cite{Kibble:1976sj}, one expects different Higgs phases in causally disconnected domains, resulting in magnetic field gradients and dumbbell formation on those scales~\cite{Vachaspati:2024vbw}. See also \cite{Patel:2023sfm,Patel:2023ybi} for an argument based on the role of dumbbells to form larger coherence lengths. On the other hand, we stress that fully realistic simulations of the cross-over nature of the EWPT~\cite{Kajantie:1996mn} including non-perturbative thermal corrections to the effective potential~\cite{Shaposhnikov:1996th}, are currently out of reach - even more so both the microphysical scale $(g T_\text{EW})^{-1}$ and the much larger cosmological scale $H_\text{EW}^{-1}$ play a role. Models and intuitions borrowed from first order (see e.g.~\cite{Stevens:2012zz,Di:2020kbw,Di:2021rtb} for simulations of this case) or second order (as employed in~\cite{Zhang:2017plw ,Vachaspati:2020blt}) phase transitions may be inaccurate in capturing the dynamical processes during the electroweak symmetry breaking,
including the formation of solitonic configurations.

In view of this we proceed with a phenomenological approach.  Motivated by our result in Eq.~\eqref{eq:h_estimate}, we parametrize the helicity fraction $\epsilon$ of the magnetic field generated at the EWPT as
\begin{align}
	\epsilon = \frac{h_c}{a^3 \lambda B^2} = \left[\frac{c \dot{\theta}}{T}\right]_\mathrm{EW} \equiv \epsilon_\mathrm{EW},
	\label{eq:helicity_fraction_EWPT}
\end{align}
where $h_c$ is the comoving magnetic helicity, $\lambda$ is the coherent scale of the magnetic field,
$c$ is a numerical coefficient which from the results of the previous section we expect to be $\mathcal{O}(0.1\,\mathchar`-\mathchar`-\,1)$ if the sourced magnetic field can be attributed to dumbbell formation and decay,
and the subscript ``EW'' indicates that the axion velocity $\dot{\theta}$ and the temperature $T$ are evaluated at the EWPT.
We note that the parameter $c$ contains two different factors: (1) the net magnetic helicity of the single EW dumbbell configuration,
and (2) the fraction of the magnetic field generated from the decay of the EW dumbbell,
relative to those generated e.g.\ by the gradient of non-solitonic Higgs configurations~\cite{Vachaspati:1991nm}.
Dedicated simulations capturing both effects are required to precisely determine the value of $c$.

Based on the results and arguments of~\cite{Vachaspati:2020blt}, we will assume an overall magnitude and coherence length of the magnetic field given by
\begin{align}
	B(T_\mathrm{EW}) \simeq \sqrt{2\times 10^{-2} \rho(T_\mathrm{EW})},
	\quad
	\lambda(T_\mathrm{EW}) \simeq \frac{1}{H(T_\mathrm{EW})},
	\label{eq:MHD_ini}
\end{align}
where $\rho(T_\mathrm{EW})$ and $H(T_\mathrm{EW})$ are the energy density and the Hubble parameter at the EWPT,
and $T_\mathrm{EW} = 140$~GeV.\footnote{
Throughout this section $\rho$ denotes the energy density, not to be confused with the radial coordinate used in the previous section.}
As discussed above, these should be considered maximal values for the magnetic field and correlation length generated in the SM electroweak cross-over. On the other hand, in many extensions of the SM, the EWPT can be first order.
This case is easier to simulate (see e.g.~\cite{Stevens:2012zz,Di:2020kbw,Di:2021rtb}) and results in comparable magnetic field strengths and correlation lengths.
It is also worth noting that several attempts have been made to obtain stable EW solitons within the SM 
in non-trivial backgrounds~\cite{Holman:1992rv,Garriga:1995fv,Nagasawa:2002at},
or beyond the SM, such as two-Higgs doublet models~\cite{Dvali:1993sg,Battye:2011jj,Eto:2018hhg,Eto:2018tnk,Eto:2019hhf,Eto:2021dca,Eto:2024xvc}. 
Given that a part of the uncertainty on the magnetic field generation from the EW strings within the SM
originates from their unstable nature, these scenarios, combined with the axionic coupling, may be particularly interesting 
for generating helical magnetic fields at the EWPT with less uncertainty.

Given these initial conditions of the magnetic field, we will discuss its cosmological evolution in the next subsection.
We focus on the location and magnitude of the magnetic field at the peak of its spectrum, characterized as above,
and do not consider its spectral shape away from the peak in the following. See~\cite{Vachaspati:2024vbw}
for a recent study on the spectral shape at large scales.

\subsection{Cosmological evolution of helical magnetic field}
\label{subsec:helical_B_evol}

To follow the cosmological evolution of the magnetic field after its generation at the EWPT,
we take a simplified approach (see e.g.~\cite{Kamada:2020bmb}) and divide the evolution into four stages:
(1) adiabatic evolution until the turbulence scale (eddy scale) reaches the coherent length of the magnetic field $\lambda$,
(2) evolution towards maximally helical configuration,
(3) inverse cascade shifting the power spectrum to larger scales while keeping maximally helical configuration,
and (4) adiabatic evolution after recombination.
In particular, epoch~(3) is only reached (before recombination) if the axion velocity at the EWPT is sizable (see Eq.~\eqref{eq:cond_maximally_helical} below),
and this epoch enhances the magnitude and coherence length of the magnetic field in the current universe.

Let us start with epoch~(1). 
First, we note that the (physical) eddy scale is estimated by the Alfven velocity $v_A$ as
\begin{align}
\label{eq:l_ed}
	\lambda_\mathrm{ed} \simeq v_A t \simeq \frac{v_A}{2H},
	\quad
	v_A \simeq \frac{B}{\sqrt{\rho + p}},
\end{align}
with $p = \rho/3$ denoting the pressure of the SM plasma.
As long as $\lambda > \lambda_\mathrm{ed}$,\footnote{
If the correlation length at the EWPT is taken to be  $(gT_\mathrm{EW})^{-1}$ instead of $H_\mathrm{EW}^{-1}$, diffusion processes will instead rapidly dampen this small-scale magnetic field, resulting in negligible surviving magnetic fields at late times.
} the magnetic field is not affected by the turbulence at the smaller scales,
and hence the (physical) coherence length and the amplitude of the magnetic field evolve adiabatically as
\begin{align}
	\mathrm{epoch}~(1):~~
	\lambda \propto a,
	\quad
	B \propto a^{-2}.
\end{align}
Since $\lambda_\mathrm{ed} \propto a^{2}$ in this regime, the eddy scale at some point catches up with the coherence length,
$\lambda = \lambda_\mathrm{ed}$ and this is the beginning of the epoch~(2), the cascade regime.
In our case, we have
\begin{align}
	\lambda(T_\mathrm{EW}) \simeq \frac{1}{H(T_\mathrm{EW})},
	\quad
	\lambda_\mathrm{ed}(T_\mathrm{EW}) \simeq \frac{0.06}{H(T_\mathrm{EW})},
\end{align}
at the EWPT, and therefore the epoch~(1) lasts for a redshift of $T_\mathrm{EW}/T \sim 16$.
Once $\lambda_\mathrm{ed}$ catches up, the turbulence affects the magnetic field and the coherence length evolves as
\begin{align}
	\lambda \simeq \lambda_\mathrm{ed},
\end{align}
which, according to Eq.~\eqref{eq:l_ed}, indicates that\footnote{
	Here we include the neutrino degree of freedom in the energy density entering in the Alfven velocity.
	Although this may not be appropriate after the neutrino decoupling,
	this treatment is sufficient for our purpose since
	this has only a minor effect on our numerical result.
}
\begin{align}
	\frac{\lambda}{B} \propto \frac{g_{*s}^{4/3}}{g_*} a^{4},
	\label{eq:turbulence_1}
\end{align}
where $a$ denotes the scale factor, and $g_*(T)$ and $g_{*s}(T)$ are the effective energy density and entropy degrees of freedom, respectively.
This expression is valid as long as the coherence length is determined by $\lambda_\mathrm{ed}$, i.e.\ during epoch~(2) and~(3), 
but does not determine the evolution of $B$ and $\lambda$ separately. If the magnetic field is maximally helical,
the conservation of magnetic helicity determines the evolution, see epoch~(3) below.
However, in general, numerical simulations and/or another conserved quantity are required to fix the evolution.
Eq.~\eqref{eq:turbulence_1} can be parameterized by a parameter $\beta$ as
\begin{align}
	\mathrm{epoch}~(2):~~
	\lambda \propto \left(\frac{g_{*s}^{4/3}}{g_*}\right)^{2/3} a^{2 - \frac{\beta+1}{\beta+3}},
	\quad
	B \propto \left(\frac{g_{*s}^{4/3}}{g_*}\right)^{-1/3} a^{-2-\frac{\beta+1}{\beta+3}}.
\end{align}
For the value of $\beta$ in epoch~(2), Ref.~\cite{Kamada:2020bmb} uses $\beta = 2$ 
while more recent study indicates 
$\beta = 3/2$~\cite{Hosking:2020wom,Zhou:2022xhk,Uchida:2022vue,Yanagihara:2023qvx}
based on the approximate conservation of the Hosking integral, leading to $B^4 \lambda^5 \propto a^{-3}$,
but we stay agnostic on the specific value.
Fortunately, in our case, the magnetic field in the current universe will not depend on the specific value of $\beta$
as long as the axion velocity is sizable 
(see Eqs.~\eqref{eq:cond_maximally_helical}\,--\,\eqref{eq:lambda_current_value} below).
With this parametrization, in epoch~(2), the helicity fraction grows as
\begin{align}
	\epsilon \propto a^{\frac{2\beta}{\beta+3}},
\end{align}
assuming the conservation of the comoving magnetic helicity.
This leads to epoch~(3) when the magnetic field reaches the maximally helical configuration, $\epsilon = 1$.
In this epoch, the conserved magnetic helicity is estimated as
\begin{align}
	h_c = a^3 \lambda B^2 = \mathrm{const},
\end{align}
and by combining it with Eq.~\eqref{eq:turbulence_1}, we obtain
\begin{align}
	\mathrm{epoch}~(3):~~
	\lambda \propto \left(\frac{g_{*s}^{4/3}}{g_*}\right)^{2/3}a^{5/3},
	\quad
	B \propto \left(\frac{g_{*s}^{4/3}}{g_*}\right)^{-1/3}a^{-7/3}.
\end{align}
In other words, the conservation of the magnetic helicity fixes $\beta = 0$ in the maximally helical case.
Finally, after the recombination, the magnetic field evolves adiabatically again, i.e.,
\begin{align}
	\mathrm{epoch}~(4):~~
	\lambda \propto a,
	\quad
	B \propto a^{-2}.
\end{align}

We denote the scale factor at the beginning of each epoch as $a = a_i$.
The transition from epoch~(1) to~(2) happens when $a = a_2$, given by
\begin{align}
	\frac{a_2}{a_1} = \frac{\lambda(T_\mathrm{EW})}{\lambda_\mathrm{ed}(T_\mathrm{EW})} \simeq 16,
\end{align}
where we ignore the small change of the effective degree of freedom.
Since $\epsilon \propto a^{\frac{2\beta}{\beta+3}}$ during epoch~(2), 
the transition from epoch~(2) to~(3) happens at
\begin{align}
	\frac{a_3}{a_2} = 
	 \epsilon_\mathrm{EW}^{-\frac{\beta+3}{2\beta}}.
\end{align}
We note that the helicity fraction is constant during the adiabatic evolution, 
i.e. $\epsilon = \epsilon_\mathrm{EW}$ at the onset of epoch~(2).
We focus on the case of reaching a maximally helical field configuration before recombination,
\begin{align}
	\frac{a_4}{a_2} > \frac{a_3}{a_2},
	~~~\mathrm{or}~~~
	\epsilon_\mathrm{EW} > \left(\frac{a_2}{a_4}\right)^{\frac{2\beta}{\beta+3}},
	\label{eq:cond_maximally_helical_beta}
\end{align}
so that the magnetic helicity generated by the axion plays a non-trivial role.
The transition from epoch~(3) to~(4) happens at recombination, $a_4/a_0 \simeq 1/1100$
where $a_0$ is the scale factor in the current universe.
By using 
\begin{align}
	 \frac{a_2}{a_4} \simeq \frac{16 g_{*s}^{1/3}(T_\mathrm{rec})T_\mathrm{rec}}{g_{*s}^{1/3}(T_2)T_\mathrm{EW}} \simeq 1\times 10^{-11},
\end{align}
where we take $g_{*s}(T_2) = 106.75$ and $g_{*s}(T_\mathrm{rec}) = 43/11$,
and setting $\beta = 3/2$ for instance, we obtain
\begin{align}
	\epsilon_\mathrm{EW} > 5\times 10^{-8}.
	\label{eq:cond_maximally_helical}
\end{align}
This lower bound becomes weaker for larger $\beta$, for instance $\epsilon_\mathrm{EW} > 2\times 10^{-9}$ for $\beta = 2$ and $\epsilon_\mathrm{EW} > 10^{-10}$ for $\beta = 5/2$.
As long as the condition~\eqref{eq:cond_maximally_helical_beta} is met (which depends on $\beta$),
the current magnetic field strength and the coherent length do not depend on $\beta$, and are given by
\begin{align}
	B_0 &= B(T_\mathrm{EW}) \times \left(\frac{g_*(T_\mathrm{rec})}{g_*(T_2)}\right)^{1/3} 
	\left(\frac{g_{*s}(T_\mathrm{rec})}{g_{*s}(T_2)}\right)^{-4/9}
	 \times \left(\frac{a_2}{a_1}\right)^{-2} \left(\frac{a_3}{a_2}\right)^{-2-\frac{\beta+1}{\beta+3}}
	\left(\frac{a_4}{a_3}\right)^{-7/3}\left(\frac{a_0}{a_4}\right)^{-2}
	\nonumber \\
	&\simeq 4 \times 10^{-13}\,\mathrm{G}\times \left(\frac{\epsilon_\mathrm{EW}}{10^{-7}}\right)^{1/3},
	\label{eq:B_current_value}
	\\
	\lambda_0 &=
	\lambda(T_\mathrm{EW}) \times \left(\frac{g_*(T_\mathrm{rec})}{g_*(T_2)}\right)^{-2/3} \left(\frac{g_{*s}(T_\mathrm{rec})}
	{g_{*s}(T_2)}\right)^{8/9}
	\times \left(\frac{a_2}{a_1}\right) \left(\frac{a_3}{a_2}\right)^{2-\frac{\beta+1}{\beta+3}}
	\left(\frac{a_4}{a_3}\right)^{5/3}\left(\frac{a_0}{a_4}\right)
	\nonumber \\
	&\simeq 20\,\mathrm{pc}\times \left(\frac{\epsilon_\mathrm{EW}}{10^{-7}}\right)^{1/3},
	\label{eq:lambda_current_value}
\end{align}
where we use the Heaviside Lorentz units and take $g_*(T_2) = 106.75$ and $g_*(T_\mathrm{rec}) = 3.36$.

\begin{figure}
\centering
\includegraphics[width = 0.5 \textwidth]{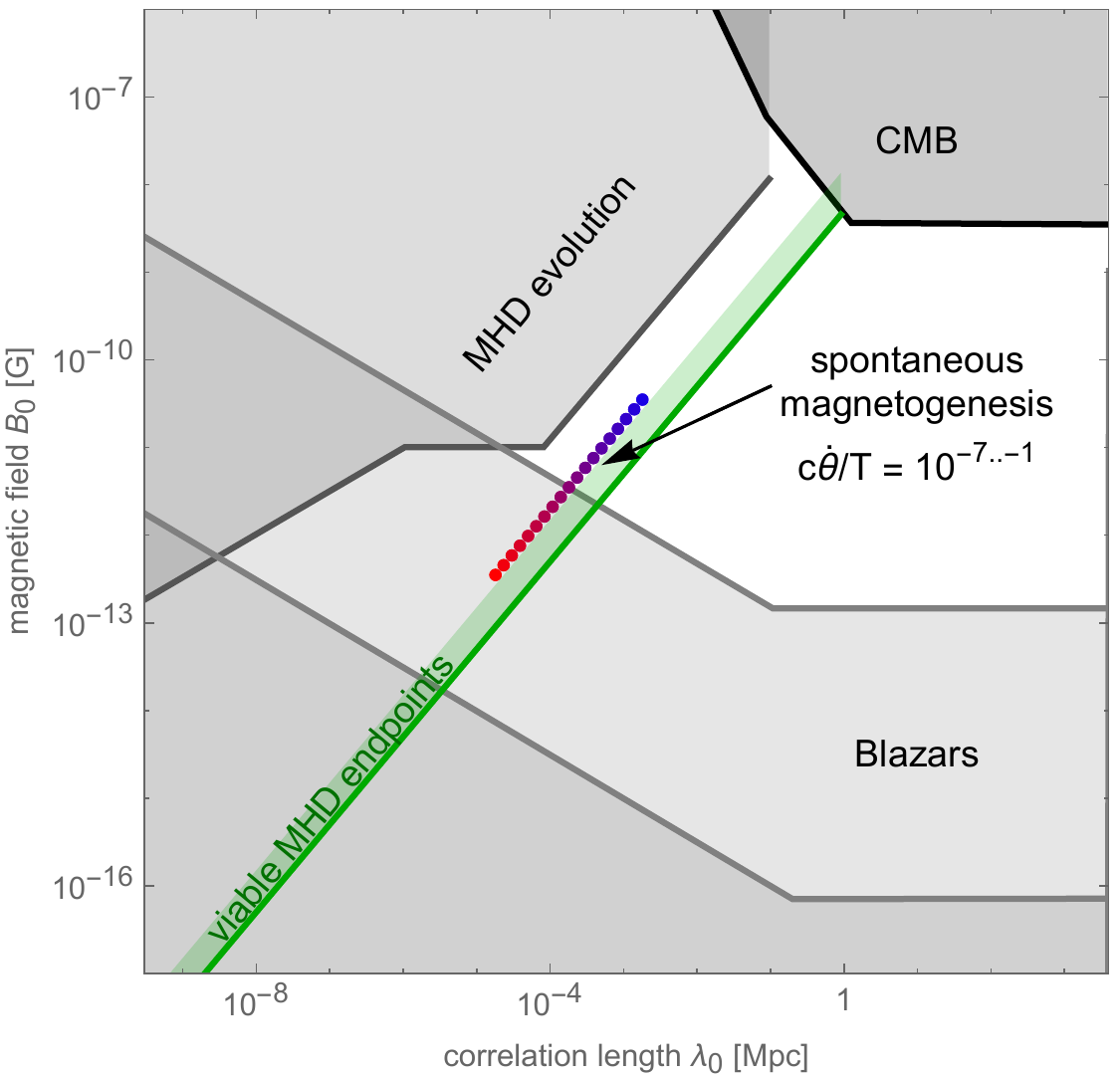}
\caption{Present day intergalactic magnetic fields. Constraints are shown in gray while viable endpoints of MHD evolution are found to the left of the green line (ranging until the theoretical ``MHD evolution'' exclusion limit), see text for details. For the lower bound on the magnetic field strength from blazar observations, we indicate both the more conservative bound~\cite{MAGIC:2022piy} as well as the more model dependent stronger bound~\cite{HESS:2023zwb}. The colored points indicate our result for $\log_{10}(\epsilon_\text{EW}) = -7$(red) .. $-1$(blue), demonstrating that spontaneous magnetogenesis driven by small axion velocities can generate magnetic fields in the right ball-park to explain the blazar observations.}
\label{fig:IGMF}
\end{figure}

The resulting present day magnetic field properties are shown in Fig.~\ref{fig:IGMF} together with cosmological and astrophysical constraints. They are consistent with the endpoints for magnetic field evolution in MHD, see in particular~\cite{Banerjee:2004df, Hosking:2022umv, Brandenburg:2024tyi}, recently summarized in \cite{Neronov:2024qtk} and depicted by the region to the left of the green line in Fig.~\ref{fig:IGMF}. See in particular~\cite{Brandenburg:2017neh} for the evolution of (partially) helical magnetic fields generated at the EWPT.
Astrophysical and cosmological constraints on IGMFs are shown in gray. The region in the top left corner (labeled ``MHD evolution'') is theoretically excluded due to the lack of a formation channel within MHD. The region to the top right is excluded by CMB observations~\cite{Planck:2015zrl,Zucca:2016iur,Calabrese:2025mza}. These are based on a number of different effects. In particular, magnetic fields increase inhomogeneities in the baryon distribution at recombination, thus altering recombination and the effective sound horizon. Stronger constraints, and even a hint for a detection have been claimed in~\cite{Jedamzik:2018itu,Jedamzik:2025cax}. These are partly driven by present tensions in cosmological data sets, and provide an interesting path towards contributing to a possible resolution. The scrutiny of these claims and data sets is beyond the scope of this work, and we show the more conservative limits only.
The lower bound on the magnetic field strength arises from the absence of GeV photons in blazar observations, attributed to the deflection of the emitted cascade of charged particles in IGMFs. Different assumptions on the blazar flux in the TeV range, i.e.\ on the primary emission, lead to more conservative~\cite{Dermer:2010mm,Taylor:2011bn,MAGIC:2022piy} or more stringent~\cite{Neronov:2010gir,HESS:2023zwb} lower bounds, which are both depicted in Fig.~\ref{fig:IGMF}.\footnote{
Note that the results for the magnetic field strength in Ref.~\cite{Zucca:2016iur} (CMB) as well as in Refs.~\cite{MAGIC:2022piy,HESS:2023zwb} (blazars) are shown in cgs units, which differ from the Lorentz Heaviside units used here (as well as in most of the MHD literature, e.g.~\cite{Durrer:2013pga,Vachaspati:2020blt}) by a factor of $\sqrt{4\pi}$. For Fig.~\ref{fig:IGMF} the constraints have been rescaled accordingly, showing predictions and constraints on the magnetic field in Heaviside Lorentz units.}

In agreement with earlier studies, see e.g.~\cite{Brandenburg:2017neh}, we find that partially helical magnetic fields generated at the EWPT can generate magnetic fields which are sufficiently strong to explain the blazar observations, though sizable uncertainties both on the interpretation of the observational limits and the magnetic field evolution remain.
The key new ingredient which we here add to this story is a mechanism which introduces a helical component to the magnetic field generated at the EWPT, due to `twisting' of the EW dumbbells in the presence of a rolling axion.
Moving forward, it would be very interesting to determine the full magnetic field spectrum generated at the EWPT in the presence of a rolling axion, and study its evolution in an MHD simulation, with the goal of reducing the theoretical uncertainty.
We further note that magnetic fields in this parameter range are of particular interest for explaining cluster magnetic fields~\cite{Banerjee:2003xk} and have also been suggested to alleviate the Hubble tension~\cite{Jedamzik:2025cax}. We refer to \cite{Vachaspati:2020blt} for a recent review on evolution, constraints and hints for cosmological magnetic fields.

\subsection{Baryogenesis}
\label{subsec:other}

A non-vanishing axion velocity during the EWPT generically leads to a generation of matter antimatter asymmetry through spontaneous baryogenesis.
The efficiency depends on the details of the axion coupling to the SM particles~\cite{Domcke:2020kcp}. In particular, if we couple the axion only to the $U(1)$ hypercharge, no baryon asymmetry is generated, since hypercharge is a conserved quantity in the SM.
Let us consider instead
the case that the axion couples to the $SU(2)$ and $U(1)$ gauge fields, and thus to $B+L$-violating operators.
In this case, assuming chemical equilibrium before sphaleron decoupling,\footnote{
Apart from the sphaleron processes, all SM processes are well in equilibrium around the EWPT. The assumption of chemical equilibrium implies that any changes to the axion velocity $\dot \theta/T$ are adiabatic and that coupling of the axion to the SM fields is sufficiently large so that the $CP$-transfer from the axion field to the SM thermal plasma is efficient. If this is not the case, then final baryon asymmetry depends on the strength of the axion coupling and has to be determined by solving a Boltzmann-type equation.}
the chemical potential of the $B+L$ asymmetry is given by
\begin{align}
	\frac{\mu_{\mathrm{B+L}}}{T_\mathrm{EW}} = \frac{144}{79}\left[\frac{\dot{\theta}}{T}\right]_\mathrm{EW}.
\end{align}
We do not generate any $B-L$ asymmetry, and hence the baryon number is given by
\begin{align}
	\eta_B = \frac{T^3}{6s}\frac{\mu_B}{T} \simeq 3\times 10^{-3}\times \left[\frac{\dot{\theta}}{T}\right]_\mathrm{EW},
	\label{eq:etaB}
\end{align}
where $s = 2\pi^2 g_* T^3/45$ with $g_* = 106.75$.
More generally, the axion may also have additional couplings to the SM fermions or to the $SU(3)$ gauge sector. A baryon asymmetry is generated whenever the direction in operator space of the axion couplings is not orthogonal to $B+L$, i.e.\ the direction of baryon number violation through sphaleron processes~\cite{Domcke:2020kcp}. In these cases, the numerical factor in Eq.~\eqref{eq:etaB} can change, but typically only by an order one factor.
Quite remarkably, a value of
\begin{align}
	\left[\frac{\dot{\theta}}{T}\right]_\mathrm{EW} \sim 10^{-7},
	\label{eq:axion_velocity}
\end{align}
gives a baryon asymmetry in the correct ballpark together with a sizeable present day intergalactic magnetic field strength.
This serendipity is only possible for values of $c$ not much smaller than unity, emphasizing the importance of a more accurate determination of this parameter based on numerical simulations.
Larger values of $(\dot \theta/T)_\textrm{EW}$ are excluded by the overproduction of baryon asymmetry, unless the axion couplings are chosen to avoid spontaneous baryogenesis~\cite{Co:2019wyp,Domcke:2020kcp,Co:2020xlh}, e.g.\ by coupling only to $U(1)$ hypercharge.

A natural mechanism to generate a sizable axion velocity, Eq.~\eqref{eq:axion_velocity}, at the EWPT
is the kinetic misalignment mechanism~\cite{Co:2019jts}.
For the QCD axion, producing the correct dark matter abundance in this framework comes with an insufficient amount of baryogenesis, while successful baryogenesis entails a (mild) overproduction of the dark matter abundance~\cite{Co:2019wyp}. This latter is in particular the case for $\dot \theta/T_\text{EW} \sim 10^{-7}$.
A solution to this is, \emph{e.g.}, that the axion velocity is damped afterwards as is the case in a heavy axion model~\cite{Co:2022aav}.

Other constraints on the $CP$-violation prior to the EW phase transition are less constraining. For $(\dot \theta/T)_\textrm{EW} \gtrsim 10^{-4}$, the axion motion can trigger the chiral plasma instability of the SM plasma, which leads to an exponential growth of helical hypermagnetic fields which in turn lead to an overproduction the baryon asymmetry at the EWPT~\cite{Domcke:2022uue}. This constraint can be evaded if the axion motion is suppressed at temperatures above the decoupling of the electron Yukawa coupling, $T \sim 10^5$~GeV. Moreover, we note that the constraints discussed in~\cite{Uchida:2024ude} do not apply since we generate the (electro)magnetic field only at the EWPT, as opposed to generating a hypermagnetic field before the EWPT.

Finally, the twisted EW dumbbell configuration itself has a non-zero baryon number as the baryon charge is related
to the Chern-Simons number through the anomaly equation, which is non-zero for the twisted configuration.
However, the EW dumbbell is unstable and decays, which changes the baryon number by the opposite amount, 
and therefore we expect that there is no net baryon number production after the formation and subsequent decay of the EW dumbbells.
Vice versa, the helicity generated through the anomaly equation from the observed baryon asymmetry is too small to have any observable consequences~\cite{Vachaspati:2020blt}.

\section{Conclusion and outlook}
\label{sec:conclusion}

Axion(-like particle) dynamics in the early universe may have rich phenomenological consequences, such as
driving cosmic inflation~\cite{Freese:1990rb,Anber:2009ua,Berghaus:2019whh}, generating the baryon asymmetry of the universe~\cite{Affleck:1984fy,Cohen:1987vi,Cohen:1988kt},
and providing an origin of dark matter~\cite{Hu:2000ke,Co:2019jts}, among others.
In this paper, we have added yet another item to the list: 
the generation of a helical magnetic field at the EWPT by a rolling axion,
which we call \emph{spontaneous magnetogenesis}.

At the EWPT, the SM Higgs and gauge bosons can form EW dumbbell configurations, i.e.\ $Z$-strings with the Nambu monopoles
attached at the endpoints~\cite{Vachaspati:1991nm,Vachaspati:2020blt}, through the Kibble mechanism.
The EW dumbbell is unstable, leaving remnant magnetic fields behind after its decay~\cite{Vachaspati:1992fi}. In the standard picture, these are erased rapidly by diffusion in the subsequent cosmic evolution.
We have shown that, in the presence of the axion coupling to the SM gauge bosons, the $CP$ violation contained in
the axion velocity induces a `twist' of the EW dumbbell configuration.
As a result, the remnant magnetic fields carry magnetic helicity.
The magnetic helicity is an approximately conserved quantity and therefore 
it prevents magnetic field diffusion during cosmic evolution~\cite{Brandenburg:2017neh}.
If the magnetic fields sourced during the SM EWPT are sufficiently strong and correlated over sizeable length scales, as indicated by recent numerical studies~\cite{Diaz-Gil:2008raf,Zhang:2019vsb,Vachaspati:2020blt}, we
find that an axion velocity of $\dot{a}/f T \sim \mathcal{O}(10^{-7})$ at the EWPT can simultaneously
explain the intergalactic magnetic field indicated by the blazar observations and the baryon asymmetry of the universe,
via the spontaneous magnetogenesis and baryogenesis mechanisms.

There are several theoretical aspects to be explored further.
First, we stress the importance of a more precise evaluation of the magnetic helicity generated at the EWPT.
The magnetic helicity is expected to be proportional to the axion velocity $\dot{\theta}$ for $\dot{\theta} \ll T_\mathrm{EW}$,
and we introduce the proportionality coefficient ``$c$'' in our estimate (see Eq.~\eqref{eq:helicity_fraction_EWPT}).
This parameter contains the information of the magnetic helicity after the decay of a single EW dumbbell,
as well as the ratio of the magnetic field associated with the EW dumbbell to the total magnetic field.
Based on some initial explorations, we assume this parameter to be of ${\cal O}(0.1 \,\mathchar`-\mathchar`-\, 1)$, but a more precise determination based on e.g.\ numerical simulations would be desirable
as its size is crucial to explain the intergalactic magnetic field and the baryon asymmetry of the universe simultaneously:
if $c$ turns out to be very small, explaining the intergalactic magnetic field would result in the overproduction of baryon asymmetry for generic axion couplings to SM particles.
There are however several challenges to such simulations: i) the dynamics of the EW crossover is very complex, requiring control over thermal corrections to the effective Higgs potential, ii) if indeed the Hubble scale is the relevant scale for the coherence length of the magnetic fields, this would requiring capturing a hierarchy of scales of $T_\text{EW}/H_\text{EW} \sim 10^{15}$, which is clearly infeasible and hence one must rely on some sort of extrapolation, and iii), introducing the coupling to the axion increases the timescale of microphysical relaxation processes, requiring longer simulation times to reach convergence.

Second, we have focused on the case that the axion couples to $SU(2)_L$ and $U(1)_Y$ with the same strength, $\theta_2 = \theta_1$, to avoid the mixing between the $Z$-boson and photon.
This is merely for simplicity, and it would be interesting to extend our analysis to a more general case of 
$\theta_2 \neq \theta_1$ (see the discussion in Sec.~\ref{subsec:nambu_monopole}).
In particular, the case $\theta_2 = 0$ could be important in connection to point above regarding baryon overproduction.
If $c \ll 1$, one possibility to avoid the baryon overproduction is to couple the axion solely to $U(1)_Y$
since this coupling alone does not generate any chiral imbalance 
in the chemical transport equation due to the absence of sphaleron configurations.

Finally, in our current study, we neglected the presence of SM fermions around the Nambu monopole,
and instead focused on one particular static solution of the equations of motion involving solely the Higgs and the gauge bosons.
However, the fermions are expected to be involved in the story, especially if $\theta_2 \neq 0$. In the absence of fermions,
a constant value of $\theta_2$ is physical and transforms a monopole to a dyon via the Witten effect~\cite{Witten:1979ey,Vachaspati:1994xe}.
Including fermions, the electric charge of the dyon is expected to be erased by the chiral fermion condensation~\cite{Callan:1982ah,Callan:1982au,Rubakov:1982fp,Hamada:2022bzn}
 since a constant $\theta_2$ is no longer physical due to the fermion chiral rotation.
We leave a more systematic study on the Nambu monopole coupling to the axion velocity, with the SM fermions included,
for future work.

\paragraph{Acknowledgements}

We thank Chiara Caprini, Ruth Durrer, Alberto Roper Pol, Misha Shaposhnikov, and Tanmay Vachaspati for discussions on primordial magnetic fields,
and Ryusuke Jinno for useful discussion on the relaxation method.

\appendix

\section{Chiral chemical potential of SM fermions}
\label{app:cme_fermion}

In this appendix, we briefly discuss the effect of fermion chiral chemical potentials on the string and monopole configurations.
This is a natural question because, in the presence of a massless fermion charged under a gauge symmetry,
the axion coupling to the gauge boson is equivalent to the axion-fermion derivative coupling.
Since physics is independent of the field redefinition, we expect that the fermion chiral chemical potential
equally generates the magnetic helicity.
Here, the chiral magnetic effect (CME)~\cite{Vilenkin:1980fu,Giovannini:1997eg,Kharzeev:2007tn,Kharzeev:2007jp,Fukushima:2008xe,Kharzeev:2009pj,Son:2012wh}
plays a crucial role.

\subsection{Chiral magnetic effect and basis independence}

For simplicity, we consider the Abelian Higgs model with a massless fermion charged under the $U(1)$ gauge symmetry.
The action is given by
\begin{align}
	S = \int d^4x \sqrt{-g}\left[
	\abs{D_\mu \phi}^2 - \lambda \left(\abs{\phi}^2 - \frac{v^2}{2}\right)^2 - \frac{1}{4}F_{\mu\nu}F^{\mu\nu}
	- \frac{\theta}{4\sqrt{-g}}F_{\mu\nu}\tilde{F}^{\mu\nu}
	+ \bar{\psi}\left(i\slashed{\nabla} - (\partial_\mu {\theta}_5) \gamma^\mu \gamma_5\right)\psi\right],
\end{align}
where $\psi$ is the fermion and 
$\nabla_\mu = \partial_\mu + \frac{1}{4}\omega_\mu^{ab}\gamma_{ab} + ig A_\mu$ with $\omega_\mu^{ab}$ the spin connection.
The anomaly equation tells us that
\begin{align}
	\partial_\mu \left(\bar{\psi}\gamma^\mu \gamma_5 \psi\right) = -\frac{g^2}{8\pi^2}F_{\mu\nu}\tilde{F}^{\mu\nu}.
\end{align}
Therefore we can shift $\theta_5$ and $\theta$ by the chiral transformation, 
and the physical combination invariant under the chiral transformation is
\begin{align}
	\bar{\theta} = \theta + \frac{g^2}{2\pi^2}\theta_5.
\end{align}
Physics should depend only on this particular combination.
To show that our string and monopole configuration indeed satisfies this property,
we may note that in the case with constant $\dot{\theta}_5$, 
the derivative coupling can be viewed as a chiral chemical potential,
\begin{align}
	-\bar{\psi} (\partial_\mu {\theta}_5) \gamma^\mu \gamma_5 \psi
	= \mu_5 \psi^\dagger \gamma_5 \psi,
	\quad
	\mu_5 = -\dot{{\theta}}_5.
\end{align}
In the presence of the chiral chemical potential, the magnetic field induces a current 
as
\begin{align}
	j^i = g\bar{\psi}\gamma^i \psi = -\frac{g^2 \mu_5}{2\pi^2}B_i = \frac{g^2 \dot{{\theta}}_5}{2\pi^2}B_i,
\end{align}
and this is called the CME~\cite{Vilenkin:1980fu,Giovannini:1997eg,Kharzeev:2007tn,Kharzeev:2007jp,Fukushima:2008xe,Kharzeev:2009pj,Son:2012wh}.
We can promote it to the Lorentz covariant form as
\begin{align}
	j^{\nu} = -\frac{g^2}{2\pi^2}\partial_\mu {\theta}_5 \tilde{F}^{\mu\nu}.
	\label{eq:CME}
\end{align}
With the CME current, the equation of motion of the gauge field is given by
\begin{align}
	0 = \frac{1}{\sqrt{-g}}\partial_\mu\left(\sqrt{-g}F^{\mu\nu}\right)
	+ \frac{1}{\sqrt{-g}}\left(\partial_\mu \theta \tilde{F}^{\mu\nu} - j^\nu\right)
	+ ig\left(\phi^* D^\nu \phi - \phi D^\nu \phi^*\right),
\end{align}
and with Eq.~\eqref{eq:CME} it reduces to
\begin{align}
	0 = \frac{1}{\sqrt{-g}}\partial_\mu\left(\sqrt{-g}F^{\mu\nu}\right)
	+ \frac{1}{\sqrt{-g}}\partial_\mu \bar{\theta} \tilde{F}^{\mu\nu}
	+ ig\left(\phi^* D^\nu \phi - \phi D^\nu \phi^*\right),
\end{align}
which indeed depends only on the physical combination.
The above argument shows that the CME is required for the field-redefinition invariance, 
and guarantees the existence of the magnetic field twist.

\subsection{Primordial asymmetry in SM fermions}

Based on the previous subsection, a natural question arises: 
Can a primordial asymmetry in the SM fermions introduce magnetic helicity to the EW solitons, even without a rolling axion?
We will see that the answer is negative; within the set of particles and interactions of the SM, the primordial asymmetry, which would be required to induce the magnetic helicity, is already erased by the SM interactions at the EWPT.

To see this, we use the formalism in~\cite{Domcke:2020kcp}.
We define chemical potentials for each SM particle as
\begin{align}
	\mu_i = (\mu_{e_1}, \mu_{e_2}, \mu_{e_3}, \mu_{L_1}, \mu_{L_2}, \mu_{L_3}, 
	\mu_{u_1}, \mu_{u_2}, \mu_{u_3}, \mu_{d_1}, \mu_{d_2}, \mu_{d_3},
	\mu_{Q_1}, \mu_{Q_2}, \mu_{Q_3}, \mu_H),
\end{align}
where the subscripts $1,2$, and $3$ indicate the generations,
and the multiplicity $g_i$ of the SM particles as
\begin{align}
	g_i = (1,1,1,2,2,2,3,3,3,3,3,3,6,6,6,4).
\end{align}
These chemical potentials are related to each other through the SM interactions in thermal equilibrium.
At the temperature around the EWPT, all the SM interactions are in equilibrium. 
The transport equation is then solved as
\begin{align}
	({M}_{Xi})
	\left(\frac{\mu_i}{T}\right)
	= \begin{pmatrix} 0 \\ c_A \end{pmatrix},
	\quad
	({M}_{Xi}) =
	\begin{pmatrix} n_i^{\hat{\alpha}} \\ g_i n_i^A \end{pmatrix},
	\label{eq:mu_eq}
\end{align}
where $c_A$ corresponds to the four conserved charges, $U(1)_Y$, $U(1)_{B-L}$, $U(1)_{L_1-L_2}$, and $U(1)_{L_2-L_3}$,
with the subscript $A$ running $Y$, $B-L$, $L_1-L_2$, and $L_2-L_3$.
The charge vectors of these four charges are given by
\begin{align}
	&n_i^{Y} = \left(-1,-1,-1,-\frac{1}{2}, -\frac{1}{2}, -\frac{1}{2}, \frac{2}{3},\frac{2}{3}, \frac{2}{3},
	-\frac{1}{3},-\frac{1}{3},-\frac{1}{3},\frac{1}{6},\frac{1}{6},\frac{1}{6},\frac{1}{2}\right),
	\\
	&n_i^{B-L} = \left(-1,-1,-1,-1,-1,-1,\frac{1}{3},\frac{1}{3},\frac{1}{3},\frac{1}{3},\frac{1}{3},\frac{1}{3},
	\frac{1}{3},\frac{1}{3},\frac{1}{3},0\right),
	\\
	&n_i^{L_1-L_2} = \left(1,-1,0,1,-1,0,0,0,0,0,0,0,0,0,0,0\right),
	\\
	&n_i^{L_1-L_2} = \left(0, 1,-1,0,1,-1,0,0,0,0,0,0,0,0,0,0\right).
\end{align}
The vector space orthogonal to these vectors is spanned by the charge vectors of the interactions, $n_i^{\hat{\alpha}}$,
and Eq.~\eqref{eq:mu_eq} states that the chemical potentials along these charge vectors are erased by the interactions.
By taking the inverse of the matrix, we obtain
\begin{align}
	\left(\frac{\mu_i}{T}\right) = (M_{Xi})^{-1}\begin{pmatrix} 0 \\ c_A \end{pmatrix}.
\end{align}

We now ask if we can have a non-zero chiral imbalance for the EW gauge bosons.
For this, we compute the chiral chemical potentials corresponding to the operators $W_{\mu\nu}\tilde{W}^{\mu\nu}$ 
and $B_{\mu\nu}\tilde{B}^{\mu\nu}$.
The former trivially vanishes since the weak sphaleron process is in equilibrium before the EWPT, 
and hence we compute the latter.
Its chemical potential is defined as~\cite{Domcke:2020kcp}
\begin{align}
	\mu_{Y_5} = \sum_{i}\epsilon_i g_i \left(n_i^{Y}\right)^2 \mu_i,
\end{align}
where $\epsilon_i = \pm$ denotes right/left-handed fermions.
We can express it as
\begin{align}
	\mu_{Y_5} = \sum_i n_i^{Y_5}\mu_i,
\end{align}
where
\begin{align}
	n_i^{Y_5} = \left(1,1,1,-\frac{1}{2},-\frac{1}{2},-\frac{1}{2},\frac{4}{3},\frac{4}{3},\frac{4}{3},
	\frac{1}{3},\frac{1}{3},\frac{1}{3},-\frac{1}{6},-\frac{1}{6},-\frac{1}{6},0\right).
\end{align}
By using the equilibrium solution, we obtain
\begin{align}
	\frac{\mu_{Y_5}}{T} = n_i^{Y_5} (M_{Xi})^{-1}\begin{pmatrix} 0 \\ c_A \end{pmatrix}
	= 0,
\end{align}
and therefore there is no chiral chemical imbalance in the SM gauge sector.
This is expected since all the conserved charges of the SM are (by definition) free from both the $SU(2)_L$ and $U(1)_Y$ anomalies.
In other words, there is no approximate symmetry broken only by the $U(1)_Y$ anomaly within the SM;
they are all either broken by other interactions or free from the $U(1)_Y$ anomaly.
Note that all the SM interactions, including the electron Yukawa interaction, are important for this conclusion.
For instance, $\mu_{Y_5} \neq 0$ is possible if the electron Yukawa interaction is out of thermal equilibrium~\cite{Domcke:2022uue,Domcke:2022kfs}.

\small
\bibliographystyle{utphys}
\bibliography{ref}
  
\end{document}